# Temperature development of glassy α-relaxation dynamics determined by broadband dielectric spectroscopy

P. Lunkenheimer*, S. Kastner, M. Köhler, and A. Loidl

*Experimental Physics V, Center for Electronic Correlations and Magnetism, University of Augsburg, 86135 Augsburg, Germany*

We present the temperature dependence of α-relaxation times of 13 glass formers determined from broadband dielectric spectroscopy, also including data from aging measurements. The data sets partly cover relaxation-time ranges of up to 16 decades enabling a critical test of the validity of model predictions. For this purpose, the data are provided for electronic download. Here we employ these results to test the applicability of the Vogel-Fulcher-Tammann equation and a recently proposed new approach that was demonstrated to provide superior fits of a vast collection of viscosity data.

Glass-forming materials are characterized by a number of unusual properties, which seem to be inherent to the glassy state of matter and which are quite universally found in such different types of materials as, e.g., glass forming alcohols, polymers, or metallic glasses [1,2,3]. Maybe the most prominent examples are the non-exponential time dependence of their relaxational behavior and the non-Arrhenius temperature dependence of their structural α-relaxation dynamics, measured, e.g., by the average relaxation times $\langle\tau\rangle(T)$ or by the viscosity $\eta$ [4]. While the details are not completely clarified yet, non-exponentiality nowadays usually is ascribed to the heterogeneous nature of the glassy and supercooled-liquid state giving rise to a distribution of relaxation times [5]. Explaining the non-Arrhenius behavior of glassy matter has proven to be a much more difficult task: There are numerous competing theoretical models of the glass transition that can describe the observed behavior with various levels of precision [1,3,6] and the glass-physics community still is far from reaching any consensus in the settlement of this question.

Thus, most experimentalists, when facing the problem how to parameterize their results on the temperature-dependent relaxation time, revert to time-honoured phenomenological approaches, the most prominent one being the Vogel-Fulcher-Tammann (VFT) law, already proposed more than 80 years ago [7]:

$$y = y_0 \exp\left[\frac{B}{T - T_{VF}}\right] \qquad (1)$$

with $y = \tau$ or $\eta$. Here $B$ is a parameter corresponding to the hindering barrier for the Arrhenius case (i.e., for $T_{VF} = 0$), and $y_0$ is the limiting high-temperature value. In Eq. (1) the non-Arrhenius behavior is taken into account by introducing the Vogel-Fulcher temperature $T_{VF}$ as an additional parameter leading to a divergence at $T = T_{VF}$. This divergence may be taken as indication of a phase-transition-like "ideal" glass transition that would occur at a temperature $T_{VF}$ below the glass temperature $T_g$, which, however, is avoided for dynamical reasons. It provides some support to theories that assume such a transition underlying the evolution of the glass state [1,6]. However, it should be noted that in a very recent work it was concluded that there is no real compelling experimental evidence for the divergence suggested by the VFT law [8]. Equation (1) nowadays usually is employed in its modified version replacing $B$ by $DT_{VF}$ [9]. The strength parameter $D$ can be taken as a measure of the deviation from Arrhenius behavior. Despite its simplicity, the VFT equation works astonishing well, especially if having in mind that it is employed to describe quantities often varying over many decades. However, especially if considering data from dielectric broadband measurements often covering $\tau$ ranges of more than 10 decades and sometimes even exceeding 15 decades, deviations from VFT behavior are frequently observed [10,11,12]. In various cases, the temperatures where such deviations are found were analyzed in detail and considered to be of importance for the understanding of the glass transition (e.g., [10,11,13,14,15]).

In a recent work [16], Mauro et al. have suggested an alternative to the VFT law, avoiding the divergence inherent to this formula, namely:

$$y = y_0 \exp\left[\frac{K}{T}\exp\left(\frac{C}{T}\right)\right] \qquad (2)$$

This equation was first proposed by Waterton in 1932 [17]. Similar but not identical relations can also be found in works by Angell and Bressel [18] and Hecksher et al. [8] where it was already pointed out that they avoid the divergence at $T_{VF}$ suggested by Eq. (1). In Ref. [16], Eq. (2) was deduced from the Adam-Gibbs equation, using constraint theory to model the temperature-dependent configurational entropy. $K$ and $C$ are related to activation energies considered in the model [16]. Eq. (2) was found to provide a superior description compared to the VFT equation when applied to a large set of viscosity data obtained on a variety of oxides and molecular liquids.

The work by Mauro et al. [16] prompted us to collect all our relaxation-time data obtained from broadband dielectric spectroscopy over the last two decades, including also so far unpublished results [19], and to test this new approach by comparing it to fits with the VFT law. The investigated materials include molecular glass formers, ionic melts, and a plastic crystal. Some of the data sets also comprise results from aging measurements [20], considerably extending the covered range of relaxation times to values up to $10^6$ s. Of course, numerous further alternatives to the VFT law have been proposed during the past decades (see, e.g., [21] for some examples) but it is impossible to discuss them all in the present work (for fits of some of the present data sets with other approaches, see [22,23]). Therefore, aside of information on the applicability of Eq. (2), the purpose of the present work is also to provide a comprehensive collection of relaxation time data extending over up to 16 decades, whose fitting may serve as a benchmark test of other approaches describing glassy dynamics. For this purpose, the present data sets are provided online for downloading [24].

*Corresponding author. Email address:
Peter.Lunkenheimer@Physik.Uni-Augsburg.de

Most of the dielectric relaxation-time data of the present work were determined from broadband spectra obtained by combining a variety of different experimental techniques as discussed in detail, e.g., in [25,26,27]. For molecular glass formers and plastic crystals, the α-relaxation leads to a peak in the dielectric loss spectra. The relaxation times were either determined by fitting the peaks obtained for different temperatures using the usually employed empirical functions as,
e.g., the Cole-Davidson function or by reading off the peak positions $\nu_p$ [22,25]. For glass forming ionic melts or ionic liquids, the loss spectra are dominated by charge transport contributions and no loss peaks are detected. Instead it is common practice to evaluate the dielectric modulus $M^* = 1/\varepsilon^*$ (with $\varepsilon^*$ the complex dielectric permittivity) to gain information on the α-relaxation process [28]. The spectra of the imaginary part $M''(\nu)$ reveal a peak, from which the relaxation time can be determined. The relaxation times obtained from fits of loss or modulus peaks can significantly depend on the employed fitting function. Assuming a disorder-induced distribution of relaxation times [5], an average relaxation time $\langle\tau\rangle$ can be calculated from the fit parameters [29] (e.g., $\langle\tau\rangle = \beta_{CD}\tau_{CD}$ for the Cole-Davidson function). This quantity is best suited to compare results from different fitting functions and thus considered in the present work. For the cases where the peak positions were read off, the quantity $1/(2\pi\nu_p)$ provides a good approximation of the average relaxation time. For most of the investigated materials, dielectric relaxation time data have been previously published by other groups, however, usually in a significantly reduced temperature range. Quite generally our data agree reasonably with literature [30]. For glycerol and propylene carbonate this was explicitely demonstrated in Ref. [25].

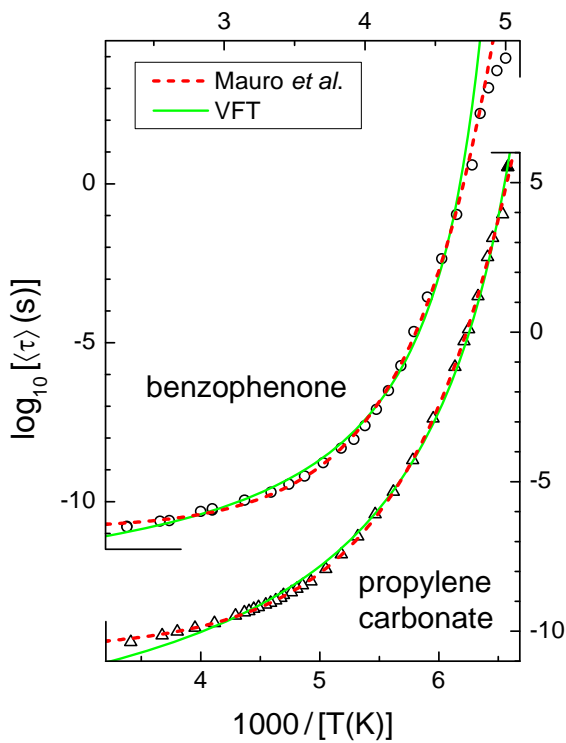

FIG. 1 (color online). Arrhenius plots of the average relaxation times of benzophenone [32] and propylene carbonate [23,25] with fits using Eqs. (1) and (2). The closed symbol denotes data from aging measurements [20]. For benzophenone only the data at $T > T_g = 212$ K have been fitted.

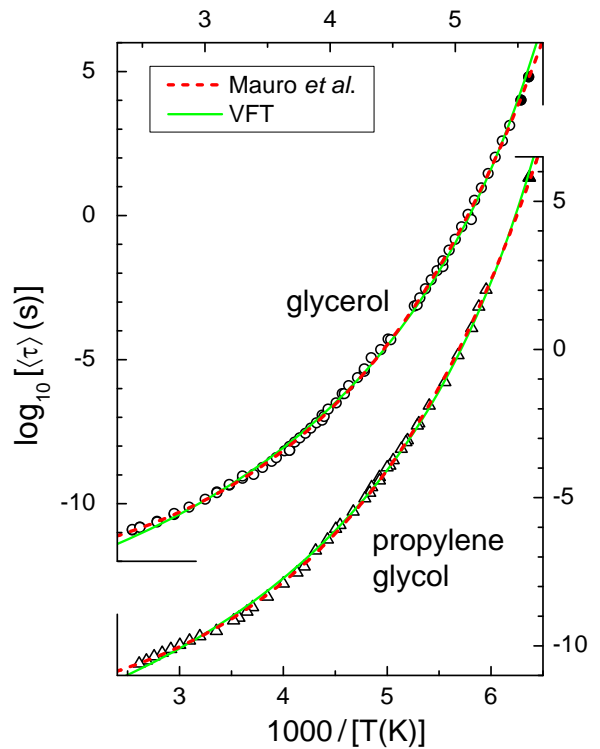

FIG. 2 (color online). Arrhenius plots of the average relaxation times of glycerol [20,25] and propylene glycol [20,34] with fits using Eqs. (1) and (2). The closed symbols denote data from aging measurements [20,35].

Table 1: Parameters of the fits of $\tau(T)$ with Eqs. (1) and (2) shown in Figs. 1-5. In addition, the glass temperature and fragility index are provided (for $m$ taken from literature references are given). The materials are ordered by increasing fragility. The last column presents the ratios of the $\chi^2$ values obtained from the fits with the two functions. A ratio smaller than unity indicates superior fits with Eq. (2).

| Material | $T_g$(K) | $m$ | VFT, Eq. (1) | | | | Mauro et al., Eq. (2) | | | | $\chi^2_M / \chi^2_{VFT}$ |
| --- | --- | --- | --- | --- | --- | --- | --- | --- | --- | --- | --- |
| | | | $\tau_0$(s) | $D$ | $T_{VF}$(K) | $\chi^2_{VFT} \times 100$ | $\tau_0$(s) | $K$(K) | $C$(K) | $\chi^2_M \times 100$ | |
| propylene glycol | 168 | 48 [34] | $1.21 \times 10^{-14}$ | 16.8 | 115 | 1.63 | $5.54 \times 10^{-13}$ | 521 | 396 | 0.704 | 0.43 |
| ethanol | 99 | 52 | $6.06 \times 10^{-11}$ | 8.15 | 76.5 | 0.268 | $2.70 \times 10^{-9}$ | 59.4 | 365 | 0.692 | 2.6 |
| glycerol | 185 | 53 [49] | $3.94 \times 10^{-15}$ | 15.8 | 132 | 2.05 | $2.29 \times 10^{-13}$ | 517 | 471 | 0.796 | 0.39 |
| Freon112 | 88 | 68 [47] | $4.24 \times 10^{-14}$ | 10.9 | 67.5 | 4.09 | $2.32 \times 10^{-12}$ | 116 | 279 | 4.67 | 1.14 |
| dipropylene glycol | 193 | 69 [34] | $2.24 \times 10^{-14}$ | 10.8 | 149 | 0.304 | $3.04 \times 10^{-12}$ | 181 | 676 | 0.538 | 1.8 |
| Salol | 218 | 73 [49] | $6.76 \times 10^{-15}$ | 7.18 | 182 | 17.4 | $5.40 \times 10^{-12}$ | 17.1 | 1301 | 5.92 | 0.34 |
| tripropylene glycol | 189 | 74 [34] | $7.14 \times 10^{-14}$ | 8.88 | 151 | 0.238 | $3.53 \times 10^{-12}$ | 155 | 685 | 0.903 | 3.8 |
| xylitol | 248 | 86 | $2.80 \times 10^{-14}$ | 6.81 | 207 | 1.07 | $5.87 \times 10^{-12}$ | 35.8 | 1320 | 5.48 | 5.1 |
| CKN | 333 | 93 [49] | $9.94 \times 10^{-16}$ | 6.69 | 273 | 7.11 | $6.30 \times 10^{-13}$ | 27.7 | 1890 | 4.13 | 0.58 |
| CRN | 333 | 100 [27] | $1.15 \times 10^{-14}$ | 4.72 | 285 | 0.882 | $4.75 \times 10^{-12}$ | 6.93 | 2320 | 0.522 | 0.59 |
| propylene carbonate | 159 | 104 [49] | $1.14 \times 10^{-13}$ | 5.85 | 134 | 5.32 | $2.45 \times 10^{-11}$ | 8.87 | 978 | 1.47 | 0.28 |
| sorbitol | 274 | 118 | $5.42 \times 10^{-14}$ | 5.17 | 233 | 1.84 | $8.55 \times 10^{-12}$ | 13.1 | 1709 | 3.19 | 1.73 |
| benzophenone | 212 | 125 [32] | $8.00 \times 10^{-13}$ | 3.32 | 191 | 4.53 | $1.45 \times 10^{-11}$ | 5.89 | 1448 | 1.20 | 0.26 |

In addition to the analysis of dielectric spectra, relaxation times at $T < T_g$ were determined from the analysis of aging measurements using a modified stretched-exponential law [20]. In Ref. [20] it was demonstrated that aging is governed by the same dynamics as the structural α-relaxation. As conventional spectroscopic measurements below $T_g$ are impracticable because of the long waiting time necessary to reach equilibrium, aging measurements are the most feasible way to extend $\tau(T)$ curves to considerably longer relaxation times than usually covered by dielectric spectroscopy. However, in ionic melts some care has to be taken because, in contrast to aging measurements, the ionic dynamics detected by dielectric spectroscopy decouples from the true structural relaxation [20,31]. Thus in those cases no data points from aging were used for the fits.

The fits of the obtained $\tau(T)$ data were performed using the least-square fitting routine of the "Origin" computer software (OriginLab Corporation, Northampton, USA). The logarithms of the relaxation times were fitted using equal weighting for all data points. The reported $\chi^2$ values, used to estimate the quality of the fits, are defined by $\chi^2 = 1/(n-p) \Sigma[\log_{10}(\tau_m)-\log_{10}(\tau_f)]^2$. Here $\tau_m$ and $\tau_f$ are the measured and calculated relaxation times, respectively. The quantity $n-p$ is the number of degrees of freedom of the fits with $n$ the number of data points and $p$ the number of fit parameters (in the present case, $p = 3$ for both equations).

Figures 1 - 4 show some of the obtained relaxation-time results in detail. Solid and dashed lines represent fit curves with Eqs. (1) and (2), respectively. The fit parameters and $\chi^2$ values are listed in Table 1. Figure 1 presents results for the low-molecular-weight glass-formers propylene carbonate [23,25] and benzophenone [32]. For benzophenone, the uppermost five data points where disregarded for the fits as they were taken at $T < T_g$ where a weaker temperature dependence is expected due to the sample falling out of equilibrium [nevertheless, the fits with Eq. (2) match some of these points quite well]. The closed symbol for propylene carbonate indicates a data point from aging, extending the $\tau(T)$ curve to achieve 16 decades. For both materials Eq. (2) clearly leads to a much better description of the experimental data than the VFT law, which is corroborated by the smaller $\chi^2_M$ in Table 1 (cf. also the ratio of $\chi^2$ values in the last column of Table 1). For propylene carbonate and various other glass formers it was shown that the VFT formula only is able to describe the experimental data if assuming a transition between different temperature regions where different laws, e.g., VFT and Arrhenius, are valid10 [10,11]. In contrast, no such transitions have to be assumed if using Eq. (2), which is able to satisfactorily describe the whole available dynamic region of 14-16 decades in these materials.

A similar statement can be made for glycerol [20,25] and propylene glycol [20,33,34,35] (Fig. 2). Compared to benzophenone and propylene carbonate, here the VFT law leads to a better, but still not perfect description of $\tau(T)$, which, including the results from aging, extends over about 16 decades. In contrast, the fits with Eq. (2) are superior, which also becomes obvious from the significantly smaller $\chi^2$ in Table 1. However, not in all cases, Eq. (2) provides a superior fit of the experimental data: In Fig. 3, the relaxation time data of another two typical molecular glass formers are shown, namely of sorbitol [35] and xylitol [20,36]. In contrast to the previous

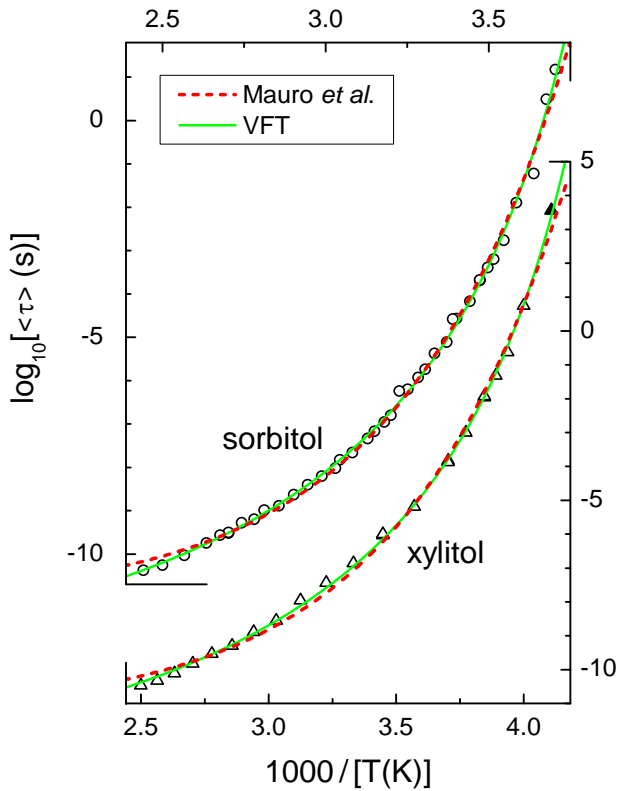

FIG. 3 (color online). Arrhenius plot of the average relaxation time of sorbitol [35] and xylitol [20,36] with fits using Eqs. (1) and (2). The closed symbol denotes data from aging measurements [20].

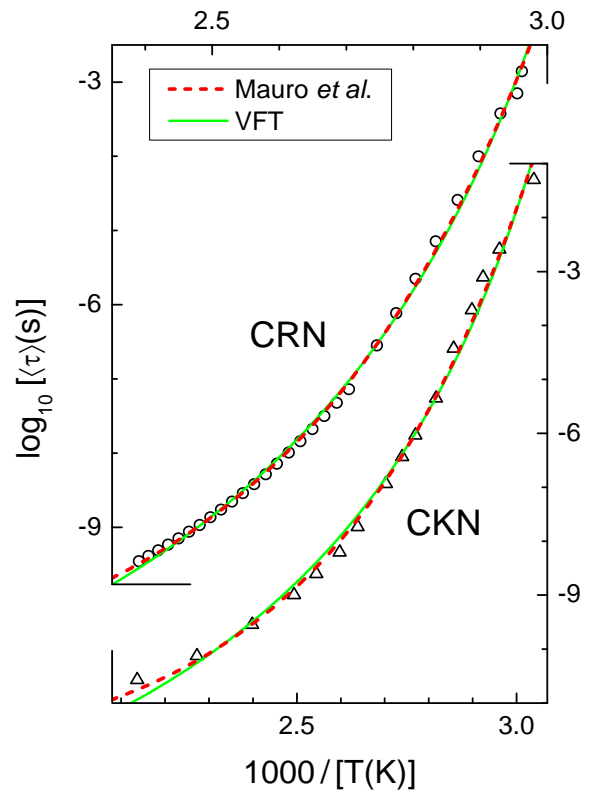

FIG. 4 (color online). Arrhenius plots of the average relaxation times of CRN [37] and CKN [31,36,37] with fits using Eqs. (1) and (2).

examples, here Eq. (1) clearly leads to a better agreement of fit curves and experimental data than Eq. (2) (see also $\chi^2$ values in Table 1).

Figure 4 shows the results for two members of another group of glass forming materials, namely the ionic melts $[Ca(NO_3)_2]_{0.4}[RbNO_3]_{0.6}$ (CRN) [37] and $[Ca(NO_3)_2]_{0.4}[KNO_3]_{0.6}$ (CKN) [22,31,36,37]. As mentioned above, these data were determined from the dielectric modulus [28] and reflect the ionic dynamics, which at low temperatures decouples from the structural α-relaxation [31]. However, this so-called conductivity relaxation-time shows all characteristics of glassy dynamics and often is described by the VFT law. For CRN, both fit curves differ only marginally, the values of $\chi^2$ suggesting a slightly better fit of Eq. (2) (Table 1). In CKN, judging by bare eye the fit with Eq. (2) seems to be somewhat superior. It should be noted that due to decoupling effects, τ(T) in these systems could not be traced up to comparably high values as in the molecular glass formers, because when the sample falls out of equilibrium at $T_g ≈ 333$ K, the conductivity relaxation time still is much shorter than the typical $τ(T_g) = 100$ s usually found in other glass formers [31]. For CKN a single point at $T < T_g$ is included in Fig. 4, which was collected after letting the sample equilibrate for ten weeks at $T = 329$ K ($1000/T = 3.04$ $K^{-1}$) [36]. It seems to suggest an Arrhenius behavior in CKN for $T < 350$ K ($1000/T > 2.86$ $K^{-1}$), which obviously for both fit functions is difficult to account for.

Finally, in Fig. 5 we provide broadband relaxation time data of another four glass formers and of a plastic crystal. For better readability, we used a $T_g$-scaled abscissa [38] and partly have additively shifted the log⟨τ⟩ values by the numbers noted in the figure. Ethanol [39] is a special case as its τ(T) exhibits a very clear transition to a weaker, Arrhenius-like temperature dependence at high temperatures. Therefore the fits were only performed for data points at $T < 190$ K. The VFT law seems to provide a somewhat better description of the data in this range. For dipropylene glycol (DPG) and tripropylene glycol (TPG) [34], the $\chi^2$ magnitudes also suggest somewhat better fits of Eq. (1). For Salol [40,41], Eq. (2) provides a clearly better, whatsoever not perfect description, especially of the strong curvature at high temperatures.

In contrast to the other materials considered in the present work, Freon112 is not a structural glass but a so-called plastic crystal. While the centers of gravity of the molecules forming a plastic crystal are translationally ordered, their orientational degrees of freedom are not. Instead the orientational dynamics in plastic crystals exhibit the typical phenomenology of glassy freezing and these materials are often considered as model systems for canonical glass formers [42,43]. However, most plastic crystals only show relatively weak deviations from Arrhenius behavior [43,44,45], i.e. they can be classified as "strong" glass formers within Angell's strong/fragile classification scheme [9]. Therefore they are not well suited

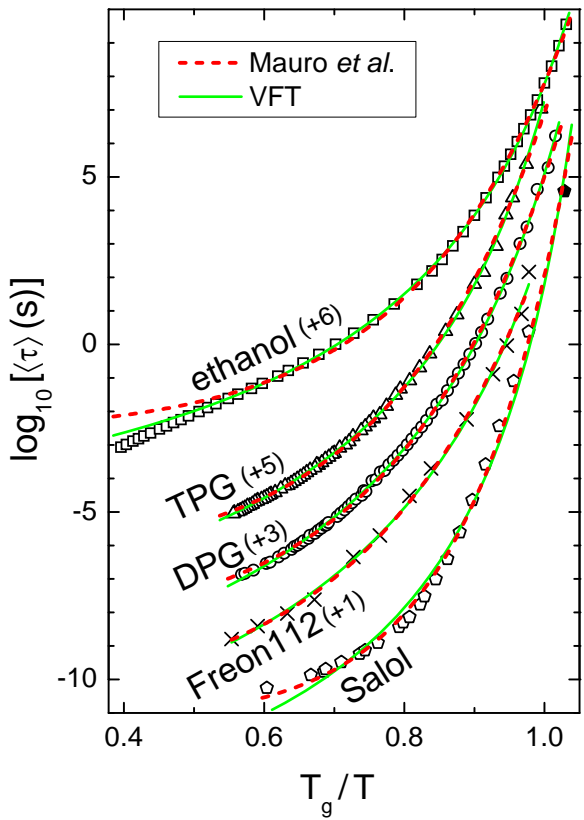

FIG. 5 (color online). Scaled Arrhenius plots of the average relaxation times of ethanol [39], tripropylene glycol [34], diproylene glycol [34], Freon112 [47], and Salol [40,41] with fits using Eqs. (1) and (2). The closed symbol denotes data from aging measurements [40]. For ethanol, only the data at $T_g/T > 0.52$ have been fitted. For better readability, some curves have been vertically shifted by the factors indicated in the figure.

for checking different models for the description of $\tau(T)$. An exception is Freon112 [(CCl$_2$F)$_2$], which belongs to the very few examples of plastic crystals with rather strong non-Arrhenius behavior [45,46,47]. As revealed by Fig. 5, the relaxation time curve of this plastic crystal can be fitted with identical quality by both Eqs. (1) and (2).

In Table 1, the parameters of all fit curves shown in Figs. 1-5 are presented, together with the results for $\chi^2$. The last column provides the ratio of the $\chi^2$ values of the two fit functions, which may serve for judging the superiority of one function in comparison to the other for the different glass formers. In seven cases, this ratio is smaller than unity and thus suggests a better description by the formula suggested by Mauro et al. and in five cases the VFT function works better (for Freon112 the ratio of 1.14 indicates a nearly equal quality). Three of the investigated glass formers were also studied in Ref. [16] (glycerol, propylene carbonate, and CKN). In the two latter cases, better fits of their viscosity with Eq. (2) could be achieved, in agreement with the findings for their relaxation times reported in the present work (cf. Table 1). For glycerol nearly equal quality of the fits with Eqs. (1) and (2) was reported in [16] but the broader dynamic range covered by the dielectric experiments of the present work allows to reveal a better quality of the fits with Eq. (2) also for this glass former.

Table 1 also provides information on the glass temperature and the fragility index $m$. The latter corresponds to the slope of the $\tau(T)$ curve in the $T_g$-scaled Arrhenius plot providing a quantitative measure of the fragility, i.e. the deviation from Arrhenius behavior [48,49]. The superiority of Eq. (1) or (2), measured by the $\chi^2$ ratio, seems neither to be correlated with $T_g$ nor with $m$ and it is not clear why for some materials one and for others the other function is better suited to describe the observed temperature dependence of the relaxation times. In [50] it was demonstrated that the VFT equation corresponds to the high-temperature limit of Eq. (2) with $T_{VF} = C$ and $B = DT_{VF} = K$. It is notable that there is a tendency for the ratios of $T_{VF}/C$ and of $DT_{VF}/K$ to be closer to unity for the less fragile glass formers (cf. parameters in Table 1). This seems reasonable because with decrasing $m$, $\tau(T)$ approaches the case of simple Arrhenius behavior, which can be equally well described by both equations.

An interesting systematic difference of the fits with both functions are the obtained limiting values of the relaxation time for infinite temperature, $\tau_0$. As revealed by Table 1, they quite generally are significantly smaller for the VFT fits than for Eq. (2). This qualitatively agrees with the results reported in [16]: Via the Maxwell relation the limiting viscosity value $\eta_0$ can be assumed to be proportional to $\tau_0$ and for the fits with Eq. (1) in most cases $\eta_0$ was found to be about 1-2 decades lower than for Eq. (2). The quantity $\tau_0$ is usually regarded as an inverse attempt frequency and assumed to be equal to a typical phonon frequency, which is of the order of 2-10 THz, implying $\tau_0$ being $2-8 \times 10^{-14}$ s. As documented in Table 1, $\tau_0$ comes closer to this region for the VFT fits. In any case, deviations from the expected range of $\tau_0$ can be explained by a transition to a different temperature dependence at very high temperatures. For ethanol (Fig. 5) this already becomes evident in the temperature region investigated in the present work as it shows a clear transition to Arrhenius behavior accompanied by very unreasonable magnitudes of $\tau_0$ for both equations (Table 1).

In summary, temperature-dependent relaxation-time data from broadband dielectric spectroscopy have been provided for 13 glass forming materials. Partly including results from aging experiments, these data sets extend over up to 16 decades and thus allow for a critical test of different model descriptions of $\tau(T)$. In the present work such tests are performed for the VFT equation and for the function recently proposed by Mauro et al. [16]. We find various examples of clear superiority of the latter but also some cases of much better fits with the VFT equation are revealed. Nevertheless, in our view it is a big success of Eq. (2) to provide a nearly perfect description of $\tau(T)$ in the classical molecular glass formers glycerol, propylene carbonate, and propylene glycol, which were investigated in a broader frequency range than any other material. Among the cases with better VFT fits, ethanol with its clear transition to Arrhenius behavior has to be regarded as special case. In DPG and TPG, the bare eye judges the deviations of both fits to be less dramatic than suggested by the $\chi^2$ ratios in Table 1. But for sorbitol and xylitol, Eq. (2) clearly fails and currently it is not clear what makes these glass formers so special. However, overall Eq. (2) seems to be a good alternative to the VFT equation, especially as in many cases it can parameterize broadband relaxation-time data with a single formula without invoking any transitions between different functions. Thus, taking into account Occam's razor, it often seems to be preferable to other approaches.

**Acknowledgements**

We thank R. Gulich for helpful discussions. This work was partly supported by the Deutsche Forschungsgemeinschaft via Research Unit FOR 1394.